# A Numerical Study of the Relationship Between Erectile Pressure and Shear Wave Speed of Corpus Cavernosa in Ultrasound Vibro-elastography

Boran Zhou[1], Landon W. Trost[2], Xiaoming Zhang[1]

[1] Department of Radiology, Mayo Clinic

[2] Department of Urology, Mayo Clinic

Correspondence:

Xiaoming Zhang, PhD

Zhang.xiaoming@mayo.edu.

Department of Radiology, Mayo Clinic, 200 1st St SW, Rochester, MN, 55905, USA.

**Abstract**

The objective of this study was to investigate the relationship between erectile pressure (EP) and shear wave speed of the corpus cavernosa obtained via a specific ultrasound vibro-elastography (UVE) technique. This study builds upon our prior investigation, in which UVE was used to evaluate the viscoelastic properties of the corpus cavernosa in the flaccid and erect states. A two-dimensional poroviscoelastic finite element model (FEM) was developed to simulate wave propagation in the penile tissue according to our experimental setup. Various levels of EP were applied to the corpus cavernosa, and the relationship between shear wave speed in the corpus cavernosa and EP was investigated. Results demonstrated non-linear, positive correlations between shear wave speeds in the corpus cavernosa and increasing EP at different vibration frequencies (100-200 Hz). These findings represent the first report of the impact of EP on shear wave speed and validates the use of UVE in the evaluation of men with erectile dysfunction. Further evaluations are warranted to determine the clinical utility of this instrument in the diagnosis and treatment of men with erectile dysfunction.

## Introduction

Erectile dysfunction (ED) is defined as the inability to achieve and/or maintain an erection satisfactory for penetrative sexual intercourse. Although prevalence estimates vary, ED occurs in approximately 52% of men aged 40-70 years and increases with age[1]. The process of erection requires the coordinated interaction of several chemical, hormonal, and mechanical pathways in the penile tissues that lead to increases in penile blood flow, expansion of the corpora, closure of venous sinuses, and ultimately tumescence [2]. As ED and cardiovascular disease share several common comorbid conditions and pathophysiologic etiologies, ED often precedes, and may even predict major adverse cardiovascular events, such as myocardial infarction or stroke (Rosen 2004 Curr Med Res Opin, Inman 2009 Mayo Clin Proc).

From a physiologic standpoint, the penis consists of two corpora cavernosa, which serve as the primary functional units leading to penile rigidity with erection. Histologically, the corpora include a combination of fibrous tissue (collagen), smooth muscle, and endothelial cells, with the ratio of these cells largely determining the erectile capacity. In the flaccid state, the cavernosal smooth muscles are contracted, which restricts arterial flow and permits engorgement of the lacunar spaces. This leads to the low penile partial pressure of oxygen (PO2 of 30-40) observed with detumescence. With sexual stimulation, nitric oxide is released from the nerves and endothelium and relaxes smooth muscle in the corpora. This increases penile arterial blood and permits expansion of the sinusoidal spaces to result in an erection. In contrast to the detumesced state, erection results in PO2 levels of 90, and intracavernous pressures exceeding 100 mmHg [3].

The pathophysiology of ED may involve several factors, including psychological, hormonal, or more commonly, changes in the ratio of collagen to smooth muscle in the penis. The increasing collagenization of the corpora results in progressive venous leak, whereby blood entering the penis subsequently drains due to inability of the penis to occlude venous outflow channels[4]. Currently, the only definitive method of assessing the penile ratio of collagen to smooth muscle is biopsy; however, given its invasive nature, this is not performed in contemporary clinical practice.

Historically, investigators have utilized penile ultrasound to aid in the diagnosis of venous leak. However, although ultrasound is excellent in detecting vascular and anatomic abnormalities, it is only able to identify end-stage disease and is limited in the ability to identify intermediary functional states. To address this gap, penile vibroelastography has been suggested as an instrument capable of evaluating the composition of corporal tissue (via measured elasticity), and as such, may provide a more distinguishing and minimally-invasive method of assessing penile corporal function.

Our team has previously published findings of a novel ultrasound vibro-elastography (UVE) technique to noninvasively characterize the biomechanical properties of superficial and deep tissues and has demonstrated its role in assessing erectile function [5-8]. Preliminary data obtained in 10 men undergoing penile U/S at the time of clinical evaluation demonstrated that UVE results correlated with direct and indirect measures of erectile function.[9]. The current study seeks to expand upon these results by focusing on the relationship between shear wave speed of the corpus cavernosa and erectile pressure (EP).

### *Numerical modeling*

Finite element modeling was developed based on the experimental setup of our previous study [9]. In that study, to generate the necessary mechanical waves, a vibratory instrument was placed perpendicular to the penile tissue and direction of wave propagation. An ultrasound probe was subsequently located 5 mm away from the vibratory instrument to capture the reflective motion of penile tissues. Three distinct wave frequencies were selected for evaluation: 100, 150, and 200 Hz. A two-dimensional FEM model was developed in ABAQUS (version 6.14-1, 3DS Inc, Waltham, MA) in this study. The Visible Human Male digital database was used to determine the gross dimensions of a two-dimensional model of the penis in its flaccid state. The model included the skin, tunica albuginea, and corpus cavernosa (Fig. 1). The initial length and diameter of the penis were 8 and 4 cm, respectively. The densities of skin and tunica albuginea were assumed to be 1000 kg/m$^3$ while the density of corpus cavernosa was assumed to be 330 kg/m$^3$. The skin and tunica albuginea were assumed to be transverse-orthotropic elastic materials, and the corpora were assumed to be an isotropic poroviscoelastic material. Mechanical properties of the tunica albuginea and skin are listed in Table 1 [10]. The void ratio of corpus cavernosa was assumed to be 0.7 [11]. The viscoelastic properties of corpus cavernosa were based on the data obtained from our prior investigations [9].

**Table 1.** Mechanical properties of tunica albuginea and skin.

| Tissue | Ex | Ey | Gxy | v |
|---|---|---|---|---|
| Tunica Albuginea | 30 kPa | 12 MPa | 4 MPa | 0.4 |
| Skin | 12.5 kPa | 0.5 MPa | 170 kPa | 0.4 |

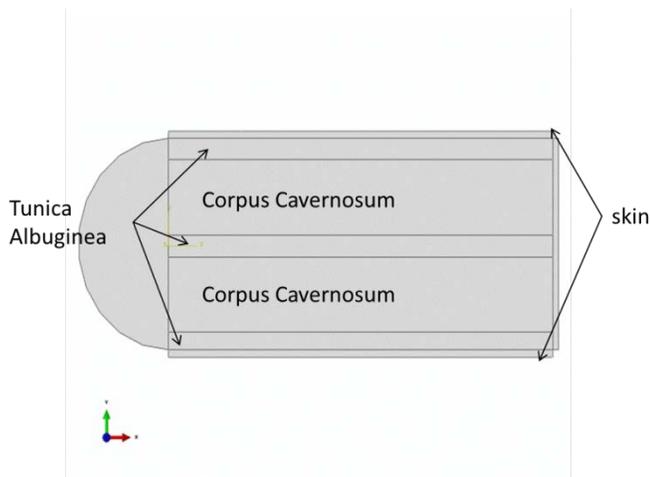

**Figure 1.** Schematic of the two-dimensional solid model of penile tissue.

The nodes on the bottom and right sides of the penile tissue were fixed for vertical and lateral movement. A uniform pressure was applied on the surfaces of corpora in the direction normal to the surfaces of porous corpus to simulate an erect state. The range of EP was set between 20 and 100 at an interval of 10 mmHg. Then, the model was excited using a segment source with the length of 3 mm on the skin, and the displacement of 0.1 mm was applied in the vertical direction at step 2. Harmonic excitations were performed at 100, 150, and 200 Hz with a duration of 0.1 s.

The mesh of skin, tunica albuginea, and corpus cavernosa were constructed using standard linear quadrilateral elements (type CPS4R), with a size of 1 mm x 1 mm used for corpus, tunica, and skin. Additional enhancements with hourglass control and reduced integration were also performed, minimizing shear locking and hourglass effects (Fig. 2). A bottom right boundary of penile tissue was associated with an infinite region to minimize wave reflections. The infinite region was meshed by infinite elements (CINPE4). Eight adjacent nodes in the corpus cavernosum 5 mm away from excitation location were horizontally selected for analyzing shear wave speed to minimize the influence of boundary effects.. The static response of penile tissue to erection was solved by the ABAQUS general static solver while the dynamic response of the tissue to the excitations was solved by the ABAQUS implicit dynamic solver with automatic step size control. Mesh convergence tests were performed so that further refining the mesh did not change the solution significantly.

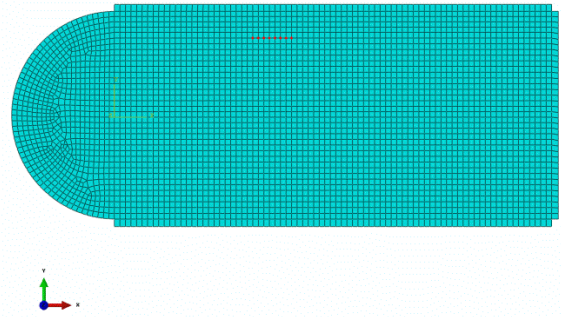

**Figure 2.** Finite element mesh of the penile tissue. Red points represent the selected nodes for analysis of shear wave speed in the corpus cavernosum.

## Results

Figure 3 is a graphical representation of the stress distribution of penile tissue during erection.

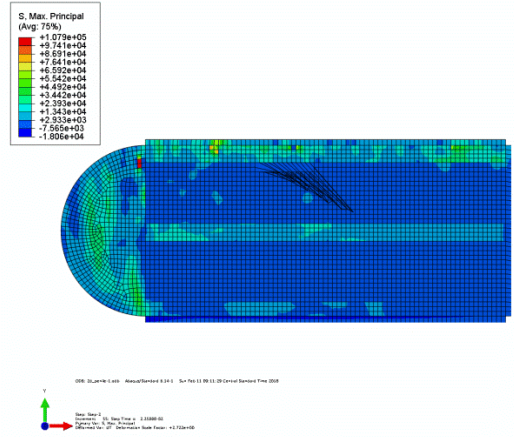

**Figure 3.** Stress distribution of penile tissue. Black lines represent displacements of eight selected nodes with wave propagation.

2D-FFT of the displacement versus time data was performed via

$$U_y(K, F) = \sum_{m=-\infty}^{+\infty} \sum_{n=-\infty}^{+\infty} u_y(x, t) e^{-j2\pi(Kmx+Fnt)}, \tag{1}$$

where $u_y(x,t)$ is the motion of the corpus perpendicular to the excitation as a function of distance from the excitation (x) and time (t). $K$ is the wave number and $F$ is the temporal frequency of the wave (Fig. 4). The coordinates of the k-space are the wave number $K$ and frequency $F$. The wave velocity was calculated as,

$$c = \frac{F}{K}, \tag{2}$$

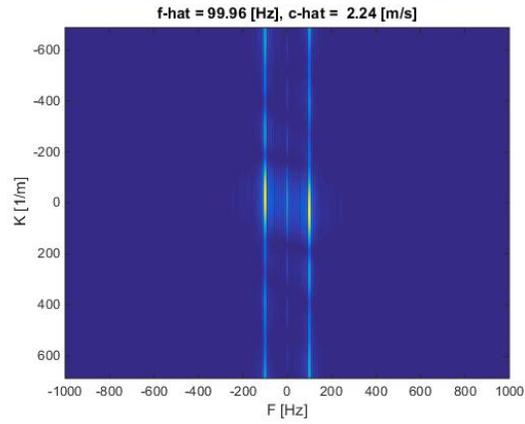

**Figure 4.** Representative k-space from 2D FFT transformation of the corpus cavernosum at 100 Hz excitation frequency.

From the numerical simulation, it showed that the wave speed in the corpus increased with excitation frequency and EP (Fig. 5).

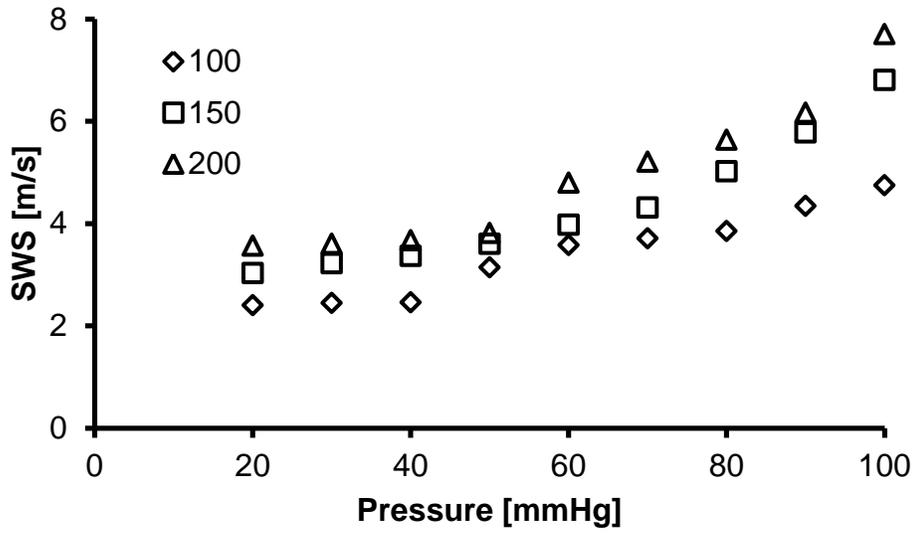
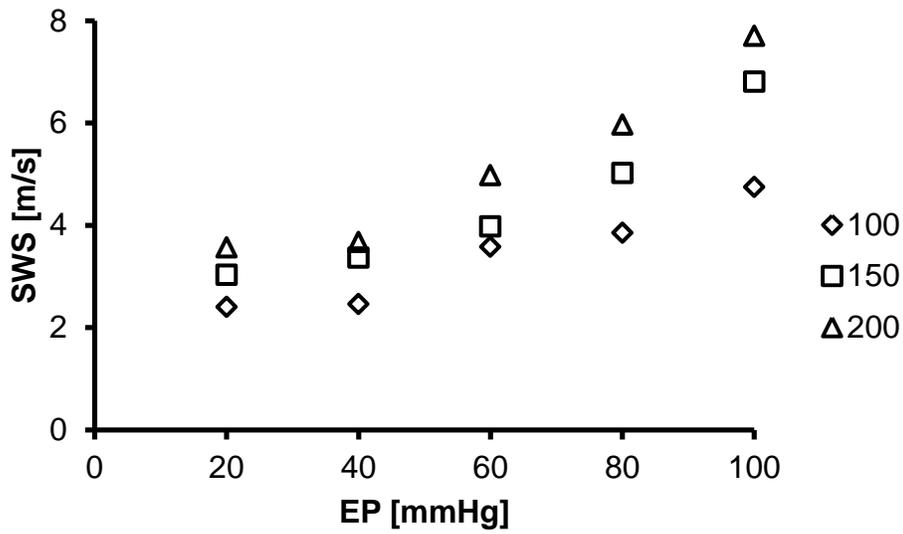

**Figure 5.** Relationship between shear wave speed (SWS) of corpus cavernosa and EP.

**Discussion**

The objective of the current study was to investigate the relationship between shear wave speed of the corpus cavernosa and EP via FEM analysis. Overall results demonstrated higher wave speeds with increasing excitation frequencies, and elevated corporal shear wave speeds with higher EP. Findings also showed that the tunica albuginea was the most dense and rigid penile tissue at erection. Importantly, results further demonstrated that the current biomechanical model was able to predict the stress distribution within different penile components, which provides added insight into underlying mechanical factors and suggests its potential role as a basic clinical tool in the diagnosing of various penile pathologies.

It is notable that several methodologic decisions and assumptions were necessary to assure the reliability of outcomes. One important decision was to utilize varying frequencies of EP (100, 150, 200 Hz) to compare a spectrum of responses, reduce variability, and normalize outcomes. Additionally, in calculating wave speeds, the corpora were assumed to be a porous material (given the known trabecular architecture), and infinite elements were assigned to the boundary of the penile tissue to minimize wave reflection. Also, as the underlying tissue microstructure have previously been shown to be similar to lung parenchyma [11], tissue density (330 kg/m$^3$) and void ratio (0.7) were extrapolated from the known parenchymal lung values and applied to the penile tissue model.

Results from this study are consistent with other published literature. In the current analysis, the stress distribution of the tunica albuginea in the erect penis was found to be ranging from 13 to 55 kPa, which is similar to the 15 kPa identified by Linder and colleagues[10]. Similarly, data on the wave speed in the corpus cavernosa were also consistent with prior publications and ranged from 2.4 (flaccid) to 7.7 m/s (erect)[9].

The concepts of tissue density and acoustic-wave propagation are not new and have been demonstrated in numerous prior studies. As a general rule, increasing tissue density results in enhanced acoustic wave-speed propagation. [12-15]. Similarly, the relationship between the echo intensity and focal tissue stress under static or dynamic loading conditions have also been shown [16, 17]. Our data further provide support for these concepts and expand upon our understanding of penile physiology. Specifically, wave speed within the corpus cavernosum is a function of intracavernosal pressure and increases in a non-linear manner with increasing EP. In the flaccid state, it is reasonable to hypothesize that the main resistance to penile expansion might be due to the cavernosal tissue itself since the tunica is more elastic and less rigid, while with increasing erectile pressure, resistive forces may be enhanced in the tunica and reduced in the corporal tissue directly.

These differences in wave speed exhibited between the flaccid and erect penis may provide enhanced ability to evaluate various pathologies within the penis. For example, studies requiring evaluation of the penile tunica albuginea (such as Peyronie's disease) may be better performed with an erect penis, while those wishing to evaluate the status of the erectile penis may be preferred in the flaccid state. Similarly, the change in wave speed between flaccid and erect states may also provide helpful information on the underlying functional status of the penis.

Enhanced diagnostic information available via the techniques described in the current and previously published manuscripts may provide practioners with useful clinical tools. The ability to distinguish between small reductions in underlying penile elasticity may signify progressive worsening of the collagen to smooth muscle ratio within the penis itself. This may help to provide a more accurate window into the underlying health of the penile tissue, a timeline as to development of erectile dysfunction, and may provide prognostic information on responsiveness to established therapies. More importantly, given the known associations between ED and underlying cardiovascular disease, this information may provide an earlier window into clinically-relevant vascular and smooth muscle alterations, which may be amendable to risk factor modification and intervention. These concepts are a natural extension of the current findings and represent a potentially exciting new area of study.

There are several notable limitations with the current study. First, multiple assumptions were required to create the current model, including the assumption that the corpus cavernosum represented a linear, homogeneous, and poro-viscoelastic material. In reality, the corpus cavernosum is far more complex, consisting of a composite of collagen fibers, smooth muscle, and endothelial cells. An improved understanding of the corpus cavernosum would likely yield a more accurate model and enhanced predictability of the effects of load-bearing constituents on the dynamic erectile response.  Second, erectile function involves the interaction of both tissue and vascular components (blood).  This complicates the analytics and may ultimately require the development of a specific fluid-structure interaction model to better understand the dynamic mechanical response exhibited with erections.  However, despite these limitations, the current study is the first of its kind to demonstrate the relationship between EP and shear wave speed.  Additionally, the study enhances our understanding of the non-linear nature of changes to elasticity and rigidity of penile tissue with erection and provides the ability to create a predictive model of response.

**Conclusion**

The current study evaluated the relationship between shear wave speed in the corpus cavernosa and EP via FEM analysis and demonstrated non-linear increases in wave speed with greater EP at frequencies ranging from 100-200 Hz. To our knowledge, this is the first study to specifically evaluate the effects of EP on the shear wave speed of the corpus cavernosa, which further validates the use of UVE for this clinical condition. This research has many potential roles in clinical practice including more accurately quantifying EP, assessing the underlying health of corporal tissue, and potentially enhancing the prognostic ability to determine future erectile function and response to specific erectogenic agents.

**Acknowledgments** - We would like to thank Mrs. Jennifer Poston for her assistance in editing the manuscript.